\documentclass[acmtog,nonacm]{acmart}

\AtBeginDocument{%
  \providecommand\BibTeX{{%
    \normalfont B\kern-0.5em{\scshape i\kern-0.25em b}\kern-0.8em\TeX}}}

\copyrightyear{2023}
\acmYear{2023}
\acmDOI{XXXXXXX.XXXXXXX}

\let\savedbaselinestretch\baselinestretch
\usepackage{listings}
\usepackage{xcolor}
\usepackage{pdfpages}
\let\baselinestretch\savedbaselinestretch

\definecolor{eclipseStrings}{RGB}{42,0.0,255}
\definecolor{eclipseKeywords}{RGB}{127,0,85}
\colorlet{numb}{magenta!60!black}

\lstdefinelanguage{json}{
    basicstyle=\normalfont\ttfamily,
    commentstyle=\color{eclipseStrings}, 
    stringstyle=\color{eclipseKeywords}, 
    numbers=left,
    numberstyle=\scriptsize,
    stepnumber=1,
    numbersep=8pt,
    showstringspaces=false,
    breaklines=true,
    frame=lines,
    string=[s]{"}{"},
    comment=[l]{:\ "},
    morecomment=[l]{:"},
    literate=
        *{0}{{{\color{numb}0}}}{1}
         {1}{{{\color{numb}1}}}{1}
         {2}{{{\color{numb}2}}}{1}
         {3}{{{\color{numb}3}}}{1}
         {4}{{{\color{numb}4}}}{1}
         {5}{{{\color{numb}5}}}{1}
         {6}{{{\color{numb}6}}}{1}
         {7}{{{\color{numb}7}}}{1}
         {8}{{{\color{numb}8}}}{1}
         {9}{{{\color{numb}9}}}{1}
}

\definecolor{GrayCodeBlock}{RGB}{241,241,241}
\definecolor{BlackText}{RGB}{110,107,94}
\definecolor{RedTypename}{RGB}{182,86,17}
\definecolor{GreenString}{RGB}{96,172,57}
\definecolor{PurpleKeyword}{RGB}{184,84,212}
\definecolor{GrayComment}{RGB}{170,170,170}
\definecolor{GoldDocumentation}{RGB}{180,165,45}

\lstdefinelanguage{rust}
{
    columns=fullflexible,
    keepspaces=true,
    frame=single,
    framesep=0pt,
    framerule=0pt,
    framexleftmargin=4pt,
    framexrightmargin=4pt,
    framextopmargin=5pt,
    framexbottommargin=3pt,
    xleftmargin=4pt,
    xrightmargin=4pt,
    breaklines=true,
    basicstyle=\ttfamily\color{BlackText},
    keywords={
        true,false,
        unsafe,async,await,move,
        use,pub,crate,super,self,mod,
        struct,enum,fn,const,static,let,mut,ref,type,impl,dyn,trait,where,as,
        break,continue,if,else,while,for,loop,match,return,yield,in
    },
    keywordstyle=\color{PurpleKeyword},
    ndkeywords={
        bool,u8,u16,u32,u64,u128,i8,i16,i32,i64,i128,char,str,
        Self,Option,Some,None,Result,Ok,Err,String,Box,Vec,Rc,Arc,Cell,RefCell,HashMap,BTreeMap,
        macro_rules
    },
    ndkeywordstyle=\color{RedTypename},
    comment=[l][\color{GrayComment}\slshape]{//},
    morecomment=[s][\color{GrayComment}\slshape]{/*}{*/},
    morecomment=[l][\color{GoldDocumentation}\slshape]{///},
    morecomment=[s][\color{GoldDocumentation}\slshape]{/*!}{*/},
    morecomment=[l][\color{GoldDocumentation}\slshape]{//!},
    morecomment=[s][\color{RedTypename}]{\#![}{]},
    morecomment=[s][\color{RedTypename}]{\#[}{]},
    stringstyle=\color{GreenString},
    string=[b]"
}

\begin{document}

\title{Synthesizing JSON Schema Transformers}

\author{Jack Stanek}
\authornote{Both authors contributed equally to this research.}
\email{jrstanek@wisc.edu}
\orcid{0009-0005-6192-2388}
\author{Daniel Killough}
\authornotemark[1]
\email{dkillough@wisc.edu}
\orcid{0009-0002-2623-0528}
\affiliation{%
  \institution{University of Wisconsin -- Madison}
  \streetaddress{1210 W. Dayton St}
  \city{Madison}
  \state{WI}
  \country{USA}
  \postcode{53715}
}

\renewcommand{\shortauthors}{Stanek and Killough}

\begin{abstract}
  JSON (JavaScript Object Notation) is a data encoding that allows structured data
to be used in a standardized and straightforward manner across systems. Schemas for
JSON-formatted data can be constructed using the JSON Schema standard, which
describes the data types, structure, and meaning of JSON-formatted data. JSON is
commonly used for storing and transmitting information such as program
configurations, web API requests and responses, or remote procedure calls; or data records,
such as healthcare information or other structured documents. 
Since JSON is a plaintext format with potentially highly complex definitions, it can be
an arduous process to change code which handles structured JSON data when its
storage or transmission schemas are modified. 
Our work describes a program
synthesis method to generate a program that accepts data conforming to a given
input JSON Schema and automatically converts it to conform to a resulting, target JSON
Schema. 

We use a top-down, type-directed approach to search for programs using a set
of rewrite rules which constrain the ways in which a schema can be modified
without unintended data loss or corruption. Once a satisfying sequence of
rewrites has been found, we pass an intermediate representation of the rewrite
sequence to a code generation backend, which synthesizes a program which
executes the data transformation. This system allows users to quickly and
efficiently modify or augment their existing systems in safe ways at their
interfaces. 
\end{abstract}



\received{14 December 2023}

\maketitle

\section{Introduction}

Industrial software development often uses JavaScript Object Notation (JSON) as a data interchange format, both when reading and writing data to disk, as well as sending and receiving data over a network. REST API endpoints, for example, frequently communicate with their clients via JSON-formatted data. The format of the JSON data can conform to a predetermined schema, which specifies the structure and types of the data that the API will send or expects to receive. The schema of some given data may not be the best format for every application; in these cases, it is usually necessary to write a large amount of tedious and error-prone data transformation code in order to wrangle the data into a more usable format. As schemas become larger and more complex, the amount of data wrangling code expands in lockstep. To simplify this process, it would be beneficial for a developer to declare a target schema that they want the data to be in and have the computer synthesize an appropriate transformer automatically. 

We propose a system that accepts two schemas in JSON Schema format, and synthesizes a unary program that accepts JSON data satisfying the input schema, and transforms it into JSON data satisfying the output schema.

In particular, we wanted to address the following three research questions:
\begin{itemize}
    \item[RQ1]: What is an efficient strategy for navigating the program search space given large and complex input and output schemas and a potentially large number of rewrite rules?
    \item[RQ2]: How can we verify that the output program correctly transforms data from one schema to another?
    \item[RQ3]: How can program synthesizers be utilized to effectively generate programs that destructively or non-destructively modify commonly used data formats?
\end{itemize}
\section{Overview}

In order to answer these research questions, some context on the problem is needed.

\subsection{What is JSON?}JSON, or JavaScript Object notation, is a standardized language that uses JavaScript syntax to represent structured data. \cite{json_spec}
JSON values can be one of several base data types, including numbers (represented as IEEE 754 floating point values), booleans, strings, and the null value.
JSON also includes constructs for nested inductive data types, such as arrays and associative arrays, dubbed "objects" as in JavaScript, which can contain values of any type.
These nested data types allow complex values to be represented in a concise, machine- and human-readable format, with a very simple language specification and implementation. An example can be seen in Listing \ref{fig:jsonFileExample}.

\hfill

\begin{lstlisting}[language=json,firstnumber=1,caption={An example JSON file describing an object with name, birth year, and paper titles properties of different types.},captionpos=b,mathescape=true,label={fig:jsonFileExample}]
{
   "name": "Alan Turing",
   "birth_year": 1912,
   "paper_titles": [ 
      "Computing Machinery and Intelligence",
      "On Computable Numbers",
      "Computability and $\lambda$-Definability"
   ]
}
\end{lstlisting}

\subsection{What is a JSON Schema?} \label{sec:whatisjsonschema}

In practice, programs expect much of the data they handle to have a particular type or structure; for instance, a web API which enumerates blog posts on a website would be expected to return an array value containing multiple objects, each of which has particular properties, such as a string representing the title, a number representing the post date, a string representing the post text, and so on.
In order to define the type and structure of JSON data, a specification for the schema of JSON values has been developed, aptly named the "JSON Schema" specification. \cite{jsonschema_spec}
Each JSON Schema characterizes the set of JSON values which conform to that schema; in some sense, a JSON Schema is a predicate on JSON values.

A JSON Schema is itself a JSON value, either a Boolean value or an object.
Boolean schemas represent trivial schemas: the schema \texttt{true} corresponds the set of all JSON values (i.e. any JSON value conforms to that schema), and the \texttt{false} schema corresponds to the empty set (i.e. no JSON value conforms to that schema.)
Non-trivial schemas are represented with JSON objects and can describe a wide variety of data types with various constraints.
In our work, we focus on a restricted subset of the JSON Schema specification in order to simplify our implementation.
Our JSON Schema specification supports defining the type of the value (one of the base or nested types) by using the \texttt{"type"} field in the schema.
This \texttt{"type"} field can specify the name of any JSON type.
Nested types, that is, arrays and objects, can have subschemas under their \texttt{"items"} and \texttt{"properties"} fields, respectively.
Array item schemas describe the JSON values that the list can contain, and object property schemas describe the values that each property in an object can have.
(Each subschema is itself a fully-fledged JSON Schema.)

It should be noted that our interpretation of the JSON Schema specification is more ``conservative'' than the official specification \cite{jsonschema_spec}, as we require that the set of properties in an object value exactly matches the set of properties defined in the schema, and that the object's properties conform to the subschemas.
The official JSON Schema specification only requires that the properties that an object value possesses must conform to the subschemas defined for that property in the top-level schema, and makes no stipulations about whether those properties must be present, nor does it limit the allowed properties to just those described in the schema.
JSON Schema allows these restrictions to be controlled with the \texttt{additionalProperties} and \texttt{required} fields, which define subschemas for properties not listed in the top-level object schema, and a list of names of required properties respectively.
In effect, our subset of JSON Schema makes two small changes to the JSON Schema's validation semantics: first, it implicitly sets \texttt{additionalProperties} to \texttt{false} for all object schemas, thus limiting the allowed properties to those specified in the schema.
Second, we implicitly set the \texttt{required} field to the list of properties specified in an object schema, which ensures that object values contain exactly the properties described in the object schema.

The JSON Schema specification also supports a number of optional constraints on values, such as upper and lower bounds for number values, maximum and minimum lengths of strings and lists, regular expressions which a string value must match, and so on.
We omit these additional constraints in our subset of the JSON Schema specification to simplify our implementation.

An example complex JSON value is shown in Listing \ref{fig:jsonschemaExample}.
The schema describes a JSON object containing information about some researcher, with properties describing a name, birth year, and a list of papers the researcher has published.

\hfill

\begin{lstlisting}[language=json,firstnumber=1,float=tp,caption={An example JSON Schema describing the set of JSON values that are objects\
        which have a ``name'' property which is a string,\ 
        a ``birth\_year'' property which is a number, \
        and a ``paper\_titles'' property which is a list of strings.},captionpos=b,label={fig:jsonschemaExample}]
{
    "type": "object",
    "properties": {
        "name": {
            "type": "string"
        },
        "birth_year": {
            "type": "number"
        },
        "paper_titles": {
            "type": "array",
            "items": {
                "type": "string"
            }
        }
    }
}
\end{lstlisting}

\subsection{Schema Transformations}

Various transformations can be defined which map values of each JSON type onto values of other types.
For instance, there exist trivial or canonical transformations between each of the base types (e.g. between numbers and strings via string formatting, between numbers and booleans by comparison to zero, and so on).
Transformations can also be mapped over values in arrays, thus ``lifting'' transformations in order to map arrays of one associated type to another.
Similarly, transformations can also be mapped over values in object properties, allowing transformations to be lifted to object types in the same way.
Object schemas admit a wider variety of transformations as well, such as deleting properties, extracting an object property, or ``annotating'' a value by inserting it into a new single-property object.
The full set of allowed transformations is enumerated in Section \ref{subsec:IR}.

Each transformation can be thought of as a function which accepts a value of the source type and returns a value of the target type. These functions represent edges in a directed graph where each vertex is a schema; source vertices represent source schemas and target vertices represent target schemas.
In this context, schema transformations can be composed through standard function composition.
The sequence of composed functions forms a path between source and target schemas, and by evaluating the composed function, a value conforming to the source schema can be converted to a value conforming to the target schema.

Alternatively, each transformation can be formulated as a rewrite rule between schemas.
For instance, each base type schema can be rewritten to any other base type schema; array schemas can be rewritten to other array schemas if the source schema's item subschema can be rewritten to the target schema's item subschema; and so on. Transformations from a type to itself, so-called ``degenerate'' transformations, are simply the identity function, and can be ignored. 

We consider base types like numbers, strings, and booleans as ``ground'' types, which indicate non-nested values that define how data is expressed in a given JSON field. Refer to Appendix \ref{sec:appendixA} for a full listing of base type transformations.

As with the function formulation above, these rewrite rules can be composed into a sequence, forming a path between valid schemas.
This path of rewrites can be realized into a program which performs the transformation on data conforming to the input schema.




\subsection{Schema Rewriting Principles} \label{subsec:principles}

Synthesized programs should behave in ways that users could can reasonably expect.
For instance, consider a transformation from an object with a numeric ``age'' property, to an object schema with a string-type ``name'' property.
In principle, since it is trivial to convert a number to a string, one could synthesize a program which simply performs this conversion and inserts the newly created string into a new object under a "name" property.
However, it is difficult to discern what this transformation ``means'' --- the meaning of the ``name'' property has changed, since the data it contains corresponds to an ``age'' value in the source.
In order to prevent such inscrutable transformations, we define a set of principles to constrain the types of composed transformations that we allow:

\begin{itemize}
    \item[1.] Schema transformations should not change the semantic meaning of data.
    \item[2.] Synthesized programs implementing a sequence of transformations should not unintentionally delete information. 
    \item[3.] The output of synthesized programs be completely determined by its input. 
\end{itemize}

These principles aim to proscribe both the introduction of arbitrary or meaningless data into the synthesized program's outputs, as well as the unintentional loss of input data.


The principles described above have a few concrete implications. 
In some sense, it is impossible for the program synthesizer to infer the exact meaning of information described by the schema, since the meaning is imposed by the semantics of programs which process the information; the only meaning that the synthesizer is nominal, i.e., the names of properties within an object schema.
Thus, in order to avoid changing the meaning of information, a synthesized program should not be allowed to modify names of properties.
This constraint is described in further detail in Section \ref{subsubsec:pathological}

Furthermore, the changes to the schema should not unintentionally delete information. Programs that we synthesize should carry over as much information as possible from the original schema, as well as satisfy the output schema, to minimize data loss in the system.
For example, if the input and output schemas describe arrays with items of particular types, the synthesized program should perform copy and transform all items in the array.
Some data loss being inevitable, however, due to some type conversions being partial---for example, converting a number to boolean will necessarily be a forgetful conversion.
Additionally, if an output object schema contains a subset of the properties in an input object schema, a synthesized program should be allowed to discard the omitted properties, since this deletion can be considered to be intentional.

Finally, any data that is described by the output schema should also be described by the input schema. If one of these conditions is not satisfied, the system should throw an error accordingly that lets the user know which condition is not being fulfilled. We don’t handle including additional fields as we don’t know what data to populate the field with, and we leave these cases up to the file owner. You can always add additional properties to files after we synthesize our output program.

\subsection{Applications and problem space}
Structured JSON data formats, such as  FHIR Health records, can frequently change as new fields become necessary for storage.
Similarly, JSON web APIs may change their data schemas over time, requiring additional server and client code to accommodate these changes; or, a developer could pull large datasets from online repositories, and need to rewrite the data format \textit{en masse} to convert them to a format more amenable to their use.
It can be an arduous process to manually write these sorts of data wrangling programs when specifications change in production systems, or when large amounts of data need to be changed at once.
Using a program synthesis approach would allow developers the ability to quickly create programs to automatically validate, filter, or clean data to a preferred specification for their work, without writing much error-prone code by hand.

\newcommand{\Alt}{\mathrel{|}}
\newcommand{\BtoB}[1][b_1\Rightarrow b_2]{\mathsf{B2B}_{#1}}
\newcommand{\Copy}{\mathsf{Copy}}
\newcommand{\PushArr}{\mathsf{PushArr}}
\newcommand{\PopArr}{\mathsf{PopArr}}
\newcommand{\PushObj}{\mathsf{PushObj}}
\newcommand{\PopObj}{\mathsf{PopObj}}
\newcommand{\PushProp}{\mathsf{PushProp}}
\newcommand{\PopProp}{\mathsf{PopProp}}
\newcommand{\ExtractProp}{\mathsf{ExtractProp}}
\newcommand{\NestObj}{\mathsf{NestObj}}
\newcommand{\InvertArr}{\mathsf{InvertArr}}
\newcommand{\InvertObj}{\mathsf{InvertObj}}

\section{System Implementation}
We have implemented a modular system which accepts input and output schemas and synthesizes a program in a target language which performs the transformation from data conforming to the input schema to a structure conforming to the output schema.
We use a type-directed top-down program search strategy in order to find a sequence of subtransformations which make up the program.

\subsection{Implementation Details}
We implemented a program synthesizer in Rust~\cite{rust}, which is structured much like a compiler, with a front-end, middle-end with intermediate representation, and modular back-end.
As input, our program accepts input and output JSON Schemas, both in JSON format, and parses them into an abstract JSON representation using the popular serde  library.~\cite{serde_rust}
The front-end is quite simple and does not contain many details of interest.
The front-end's main usefulness, other than parsing the schemas, is in producing useful error messages: for instance, the user will be notified if a schema includes an invalid type, or is missing required fields such as ``items'' for array schemas or ``properties'' for object schemas.

The middle-end of the synthesizer performs the bulk of the program search, and upon finding a rewrite path between the input and output schemas, extracts the path in an intermediate representation.
The back-end uses this intermediate path representation in order to generate code in a target language.
Thanks to our modular architecture, it is possible to implement code generation back-ends for any language; we have implemented a JavaScript back-end in our prototype due to its native support for JSON manipulation.

\subsection{Intermediate Representation} \label{subsec:IR}
The middle-end of the program synthesizer performs the program search and generates an intermediate representation which encodes the path between the schemas.
Program search is carried out in a top-down, type-directed manner.
This is done by performing a case analysis on the input and output schemas.
A description of the intermediate representation language is given in Figure \ref{fig:irsyntax}.
The nonterminal $S$ represents a complete rewrite sequence describing a path between two schemas, or a composition of multiple transformations.
The nonterminal $R$ represents either a single rewrite on a value, or a subsequence of transformations to carry out on the items of an array or the properties of an object.
The nonterminal $K$ represents a sequence of rewrites to be carried out on an the value of a property in an object.

\subsubsection{IR Grammar and Rewrite Rules}
The IR language represents both single transformations on JSON values and a rewrite rules on schemas.
The $\BtoB$ terminal represents a transformation of a single value of base type $b1$ to base type $b2$.
The $\Copy$ terminal represents a no-op, or the identity function --- it simply copies a value without modifying it. ($\Copy$ encompasses all degenerate transformations.)


For nested types, such as arrays and objects, transformation paths are more complex.
The $\PushArr$ and $\PopArr$ terminals can surround a sequence of composed transformations which map over the elements of an array; in effect, they lift a series of schema rewrites to the item subschema of an array.
The $\PushProp_p$ and $\PopProp$ terminals work similarly, but for properties in an object--- they lift the series of rewrites to the subschema of property $p$ in an object schema.
Multiple rewrite sequences wrapped with $\PushProp_p$ and $\PopProp$ must themselves be wrapped in a pair of $\PushObj$ and $\PopObj$ terminals.
Properties are allowed to be deleted from objects if the target schema's properties are a subset of the input schema's properties, since it is assumed that discarding this information is intentional; this transformation can be achieved by omitting a transformation for that key in the rewrite sequence.

More rewrites are possible for object schemas.
For instance, any property schema can be extracted from an object schema with one or more property subschemas.
Dually, a schema can be nested into a new single-property object schema under an arbitrarily-named property.
These two transformations are represented by the $\ExtractProp_p$ and $\NestObj_p$ terminals, respectively, which perform these transformations targeting the property $p$.
The most involved transformations are represented by the $\InvertArr$ and $\InvertObj$ terminals; these transformations can convert a list containing objects with some set of properties, to an object with the same properties but with its associated property subschemas transformed into arrays with corresponding item schemas, and vice-versa.

\begin{figure}
    \centering
    \begin{align*}
        R &\Coloneqq \BtoB\\
          &\Alt \Copy\\
          &\Alt \PushArr\ S\ \PopArr\\
          &\Alt \PushObj\ K\ \PopObj\\
          &\Alt \ExtractProp_p\\
          &\Alt \NestObj_p\\
          &\Alt \InvertArr\\
          &\Alt \InvertObj\\
        K &\Coloneqq \PushProp_p\ S\ \PopProp\\
        S &\Coloneqq \epsilon\\
          &\Alt R\ S
    \end{align*}
    \caption{Grammar of intermediate representation language for encoding the path between two schemas.}
    \label{fig:irsyntax}
\end{figure}

\subsubsection{Pathological rewrite sequences} \label{subsubsec:pathological}

Each rewrite rule described in the previous section individually conform to the principles described in Section \ref{subsec:principles} which enforce ``safe'' rewrites.
However, since compositions are able to be freely composed, there are certain combinations of rewrites which can in fact violate these principles.
In particular, the safe rewriting principles proscribe changing the names of properties in object schemas, as this would amount to ``changing the semantic meaning'' of the value.
Furthermore, changing the names of properties in object schemas can in effect make the rewrite path search non-deterministic for object schemas which delete one or more properties.

Consider for example an object schema with two numeric properties \texttt{"A"} and \texttt{"B"}, which we wish to rewrite to an object schema with a string property \texttt{"A"}.
If renaming properties is allowed, then two paths exist between the two schemas: either B could be deleted from the object schema, or A could be deleted and B renamed to A.
Since both paths relate the two schemas, some mechanism would be needed to select a unique path in order for the path search to be deterministic. In lieu of such a mechanism, we chose instead to prohibit renaming properties to prevent such situations. More information on experimentation can be found in Section \ref{sec:exp}.

Unfortunately, the $\ExtractProp_p$ and $\NestObj_p$ rewrites can be used in combination to effectively rename properties in object schemas, simply by sequencing the two.
Consider the object schema from before, with numeric properties A and B.
By performing the rewrites $\ExtractProp_A,\ \BtoB[\textrm{number}\Rightarrow \textrm{string}],\ \NestObj_B$, we have in effect renamed A to B as part of the transformation.

In order to prevent such abuse of these rewrite rules, we forbid rewrite sequences of the form $\ExtractProp_{p_1}\ S\ \NestObj_{p_2}$ when $p_1 = p_2$.
Note that this restriction makes the grammar of the intermediate representation non-context-free.

\subsection{Path Search}
The middle-end constructs a path between schemas by conducting a top-down type-directed program search to determine an optimal sequence of rewrites for our system.
Program search is initiated by comparing the input schema with the output schema.
If both schemas describe base types, a single-rewrite sequence is produced containing only a $\BtoB$ rewrite, where $b1$ and $b2$ are the corresponding base types.
If both schemas describe array types, the item subschemas are compared recursively, and the path between the two schemas is simply the path between the subschemas wrapped between $\PushArr$ and $\PopArr$ terminals.
If at least one of the schemas describe object types, path search is more complex.
First, if the source schema describes an object type, a search is performed to check if a path exists from a unique property subschema in the source schema, $p$, to the target schema, in which case the path is simply an $\ExtractProp_p$ rewrite followed by the path between the property subschema and the target schema.
Second, if the \textit{target} schema describes an object type, a search is performed in the opposite direction, checking if a path exists from a unique property subschema $p$ in the \textit{target} schema to the source schema.
In this case, the path is $\NestObj_p$ followed by the reversed path from the target property subschema to the source schema.
Otherwise, the property name-schema tuples are compared pairwise between the source and target schemas; the property names in the target schema must be a subset of the property names in the source schema.
Paths are searched recursively between each corresponding property subschema in the source and target schemas.
Each sub-path between property subschemas is wrapped in a pair of $\PushProp_p$ and $\PopProp_p$ terminals, and the resulting sequences are themselves wrapped in a single pair of $\PushObj$ and $\PopObj$ terminals.

\subsection{Code Generation}
Code generation is carried out by a modular back-end.
A back-end implementation for a given target language accepts a string in the intermediate representation language and produces a program in the target language.
In our implementation, we selected JavaScript as our target language, but in principle back-ends for any language could be implemented.
Our code generation algorithm is quite simple, and constructs programs in a purely syntactic manner, that is, without constructing an abstract representation of the target language such as an AST.
Generation begins by instantiating a boilerplate function wrapper into which the synthesized code can be inserted.
The function wrapper is simply an anonymous function declaration which can accept one argument, and a corresponding ``return'' statement; these two components wrap the synthesized code.
For each $\BtoB$ or $\Copy$ transformation in the intermediate representation, a corresponding JavaScript statement is emitted which carries out the transformation.
(For $Copy$ transformations, this amounts to simple assignments, and for $\BtoB$ transformations, the right hand side of the assignment is wrapped in an appropriate conversion operation.)
When a $\PushArr$ instruction is processed, a new array initialization is emitted into the program; similarly, with the $\PushObj$ instruction, a new object initialization is emitted.
Array and object variable names will be recorded on a stack which tracks the dotted paths for assignments in nested objects.
For transformation sequences surrounded by $\PushArr$ and $\PopArr$, an associated index variable will be pushed onto the stack as well when the loop template is emitted.
The lifted loop transformations will each have an assignment from the input array's indexed value to the target array's indexed value.
For $\PushProp$ instructions encountered after a $\PushObj$ instruction, the associated key will be pushed onto the variable stack and subsequent transformations will produce assignments targeting that property.
For instance, if the following sequence is encountered: 

\begin{align*}
    \PushObj\\
    \PushProp_{A}\\
    \Copy\\
    \PopProp\\
    \PopObj\\
\end{align*}

Then a new unique variable name will be pushed onto the stack when $\PushObj$ is processed, followed by the property name A, followed then an assignment from the input's A property to the unique stack variable's A property.
The $\PopProp$ instruction has the effect of popping the property name off of the stack, and the final $\PopObj$ pops the unique variable name off of the stack and emits an assignment from the stack variable to the new variable at the top of the stack (or the function's return value if the stack is empty.) 
We first need to instantiate a new wrapper array or object to contain all future transformations (until the nested type is closed), then descend one `level' until we're done populating the nested type with conversions. In essence, we push the contents of the array or object onto a stack, convert its internal properties independently, then pop back to the previous level to save its structure.
More specifically, for arrays, we create new target arrays to insert transformed values into, and in objects, we create new target objects one nested level at a time. 
We then generate our code modularly based on the determined rewrite sequence to generate a fully synthesized program.
The system starts with boilerplate code that generates a unary function that takes in a JSON file as input and returns an output file. However, the code generated in between is dependent on the results of a path found from a set of valid, `best' intermediate representation transformations. 

This method of code generation produces code reminiscent of static single assignment, where we generate a single variable for each transformation, e.g. a new array or object variable for each lifted transformation sequence.
This process does add some bloat, since a new variable is created for each transformation, but in return the implementation is greatly simplified.
An improved implementation could instead modify an AST directly instead of simply emitting target language syntax, but is beyond the scope of this project.
\section{Experiments}
\label{sec:exp}
To test our system, we have devised a set of unit tests to validate individual sections of our code.  
Unit testing ensures that particular functions appropriately follow the described behavior as well as give us an idea of how our system formats its output. These unit tests perform individual operations on files and are expected to generate small code snippets that may or may not run on their own. 
We have also conducted end-to-end tests which synthesize full programs, taking a single input file and performing the appropriate transformations as generated by our system. 

Tests were written in Rust at the bottom of each source file containing functions that modified system behavior. Each test \textit{section} was flagged with the Rust \lstinline[language=rust]{#[cfg(test)]} header and each test was flagged with the Rust \lstinline[language=rust]{#[test]} header. All tests were run using \textit{\lstinline{cargo run test}} on the command line.

This testing procedure ensures that we properly convert from a given input schema to proper intermediate data representations and, ultimately, code. Some examples of our unit tests can be found in Appendix B, Listing \ref{figB}.

We want to ensure best order performance for our output methods, so we evaluate over a series of 3 schemas that vary in length to ensure the lowest time. We should likely take an input cost function to evaluate a more objective ``best'' measurement; however, for this implementation, we do not accept an arbitrary function for a parameter and have hard-coded this evaluation into our program. Some ideas for future work in this area can be found in Section \ref{sec:futurework}. 
\section{Limitations}
Current limitations of our system note that some transformations are not fully implemented.
For instance, the $\InvertArr$ and $\InvertArr$ rewrite rules are not implemented in the back-end due to the time constraints of this project and the complexity of implementation within the architecture of our JavaScript backend.
In addition, since our system only supports a restricted subset of the JSON Schema specification, many features of JSON Schemas widely used in practice are not supported.
Some examples are constraints on string lengths, number intervals, or regular expression matching, as well as support for re-using common subschemas referentially--- to illustrate, one could ``factor out'' a schema for describing a date using a string matched against a regular expression, and re-use this common subschema across many locations in a larger schema. 
Finally, as described in \ref{sec:whatisjsonschema}, the version of the JSON Schema specification that we support is also ``conservative'' in the sense that it assumes all listed properties are implicitly required, and that no additional properties are allowed.
By relaxing this restriction, the set of JSON values that a given schema can characterize can grow, thus growing the rewrite sequence search space, and  necessitating a large amount of boilerplate run-time checks in the synthesized program.

\section{Future Work}
\label{sec:futurework}

Some ideas for future work on this project include adding the ability to specify additional invariants over the transformations and supporting additional properties or features that JSON currently supports, as enumerated in the previous section.

We have also considered adapting this project to be a plugin for applications like visual studio code, or making a containerized web application or other graphical user interface to increase adoption.

To determine a more objective ``best'' determination between the two paths, we consider choosing to accept an arbitrary cost function for each parameter in our input function, and based on its behavior, handle schema transformations accordingly. For example, we could support the addition of new properties by accepting an input function that flags all new properties and initializes them with a default value based on its type. We may also consider allowing deletions under certain conditions as expressed by this function as well, though particular behavior should be explicitly defined in documentation for the system specification as well as follow a particular format. We have not included the framework to support such a system in this project. 
\section{Conclusion}
Overall we have created a program synthesizer that generates code to effectively one-way convert two JSON schemas from one to another. 
We hope this system enables developers to more efficiently convert their data, making specifications changes less arduous.

\section{Acknowledgments}

Special thanks to Dr. Thomas W. Reps for his advice and guidance during COMP SCI 703 at UW-Madison. 

 \bibliographystyle{ACM-Reference-Format}
 \bibliography{mainbiblio}


\begin{thebibliography}{4}


\ifx \showCODEN    \undefined \def \showCODEN     #1{\unskip}     \fi
\ifx \showDOI      \undefined \def \showDOI       #1{#1}\fi
\ifx \showISBNx    \undefined \def \showISBNx     #1{\unskip}     \fi
\ifx \showISBNxiii \undefined \def \showISBNxiii  #1{\unskip}     \fi
\ifx \showISSN     \undefined \def \showISSN      #1{\unskip}     \fi
\ifx \showLCCN     \undefined \def \showLCCN      #1{\unskip}     \fi
\ifx \shownote     \undefined \def \shownote      #1{#1}          \fi
\ifx \showarticletitle \undefined \def \showarticletitle #1{#1}   \fi
\ifx \showURL      \undefined \def \showURL       {\relax}        \fi
\providecommand\bibfield[2]{#2}
\providecommand\bibinfo[2]{#2}
\providecommand\natexlab[1]{#1}
\providecommand\showeprint[2][]{arXiv:#2}

\bibitem[jso({[n.\,d.]})]%
        {json_spec}
 \bibinfo{year}{[n.\,d.]}\natexlab{}.
\newblock
\newblock
\urldef\tempurl%
\url{https://www.ecma-international.org/wp-content/uploads/ECMA-404_2nd_edition_december_2017.pdf}
\showURL{%
\tempurl}


\bibitem[dtolnay(2023)]%
        {serde_rust}
\bibfield{author}{\bibinfo{person}{dtolnay}.} \bibinfo{year}{2023}\natexlab{}.
\newblock \bibinfo{title}{Crate serde\_json}.
\newblock
\newblock
\urldef\tempurl%
\url{https://docs.rs/serde\_json/latest/serde\_json/}
\showURL{%
Retrieved 2023 from \tempurl}
\newblock
\shownote{Serde JSON Rust Library Documentation}.


\bibitem[Rust({[n.\,d.]})]%
        {rust}
Rust \bibinfo{year}{[n.\,d.]}\natexlab{}.
\newblock
\newblock
\urldef\tempurl%
\url{https://www.rust-lang.org/}
\showURL{%
\tempurl}
\newblock
\shownote{Rust Language.}.


\bibitem[Wright et~al\mbox{.}(2022)]%
        {jsonschema_spec}
\bibfield{author}{\bibinfo{person}{Austin Wright}, \bibinfo{person}{Henry Andrews}, \bibinfo{person}{Ben Hutton}, {and} \bibinfo{person}{Greg Dennis}.} \bibinfo{year}{2022}\natexlab{}.
\newblock \bibinfo{booktitle}{\emph{{{JSON Schema}}: {{A Media Type}} for {{Describing JSON Documents}}}}.
\newblock \bibinfo{type}{{T}echnical {R}eport}. \bibinfo{institution}{{Internet Engineering Task Force}}.
\newblock


\end{thebibliography}

 \appendix
    \newpage
\counterwithin{figure}{section}
\counterwithin{lstlisting}{section}

\section{Enumerating Ground to ground transformation behavior.}
\label{sec:appendixA}
\begin{figure}[!htb]
    \centering
    \includegraphics[
  width=30cm,
  height=15cm,
  keepaspectratio,
]{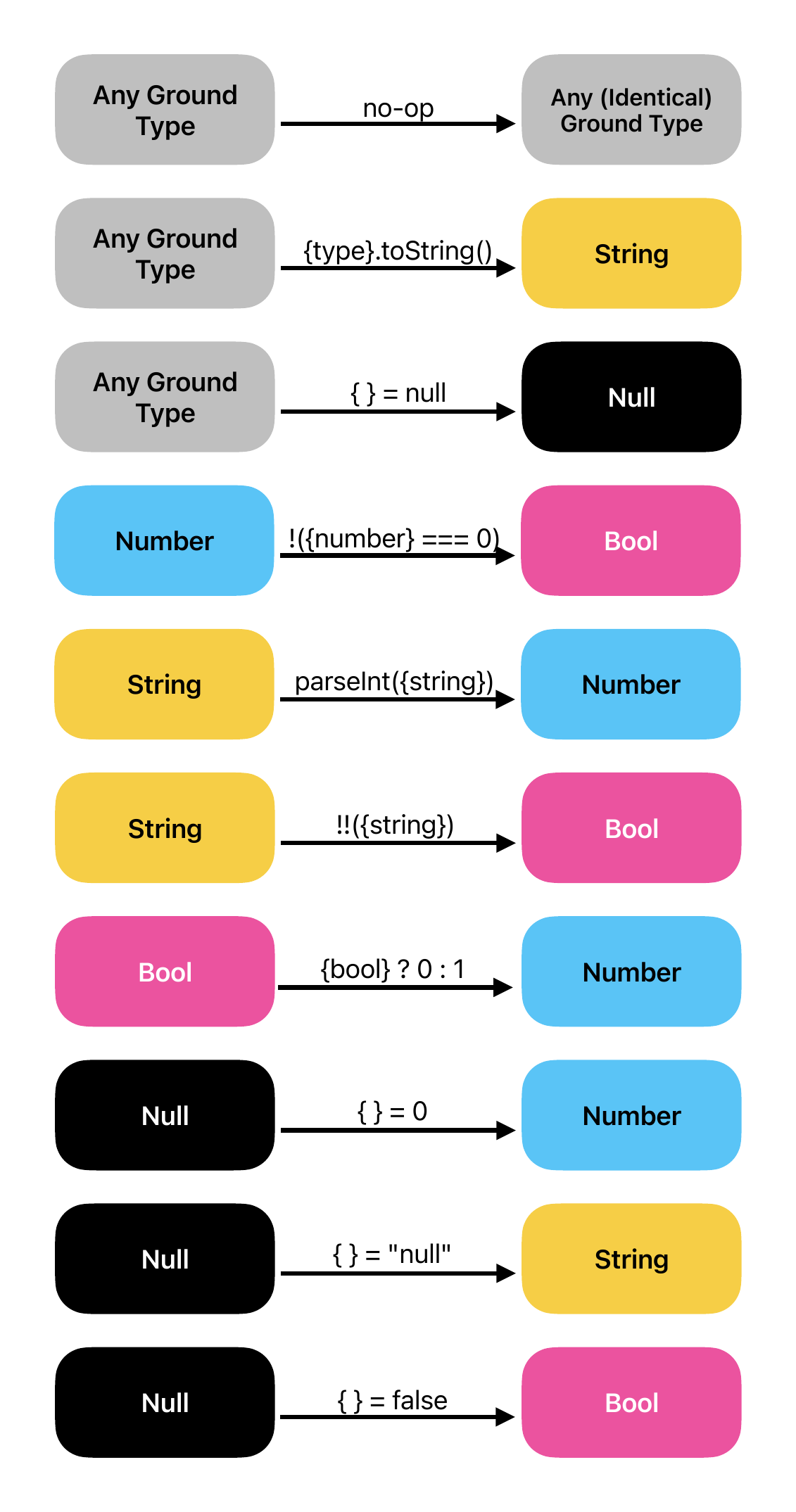}
    \caption{Ground to Ground Transformation Behavior. Types on the left are input types and types on the right are the resulting type.}
    \label{figA}
\end{figure}

\newpage
\section{Testing.}
\subsection{Unit Test Examples}
\begin{lstlisting}[language=rust, showstringspaces=false,captionpos=b,caption={Unit test demonstrating ability to use internal stack for arrays, as well as converting the type of the array (all elements within) from ground type (e.g. string) to ground type (e.g. number)},label={figB}]
#[test]
fn test_push_arr() {
    let code = JSCodegen::new("input", "output")
         .generate(vec![PushArr, G2G(Ground::String, Ground::Num), PopArr].into_iter());
    assert_eq!(
        code,
        "\
function(input) {
    let arr0 = [];
    for (let idx1 = 0; idx1 < input.length; idx1++) {
        arr0[idx1] = parseInt(input[idx1]);
    }
    output = arr0;
    return output;
}"
    )
}
\end{lstlisting}







\end{document}